\documentclass[10pt,twocolumn]{article} 

\usepackage{ol2}
\usepackage{amsmath}
\usepackage{graphicx}

\begin{document}
\thispagestyle{empty}

\twocolumn[ 

\title{Soliton transverse instabilities in nonlocal nonlinear media}

\author{YuanYao Lin$^{1,*}$, Ray-Kuang Lee$^1$, and Yuri S. Kivshar$^2$}

\address{$^1$Institute of Photonics, National Tsing-Hua University,\\ 
101,Section 2, Kuang-Fu Road, Hsinchu City 300, Taiwan\\
$^2$Nonlinear Physics Center, Research School of Physical Sciences and Engineering, \\
The Australian National University,Canberra, ACT 0200, Australia\\
$^*$Corresponding author: d928103@oz.nthu.edu.tw}

\begin{abstract}
We analyze the transverse instabilities of spatial bright solitons
in nonlocal nonlinear media, both analytically and numerically. We
demonstrate that the nonlocal nonlinear response leads to a dramatic
suppression of the transverse instability of the soliton stripes,
and we derive the asymptotic expressions for the instability growth
rate in both short- and long-wave approximations.
\end{abstract}

\ocis{190.3270 190.4420 190.6135 }
] 

Symmetry-breaking instabilities have been studied in different areas
of physics, since they provide a simple means to observe the
manifestation of strongly nonlinear effects in nature. One example
is the transverse instabilities of spatial optical
solitons~\cite{YSK00} associated with the growth of transverse
modulations of quasi-one-dimensional bright and dark soliton stripes
in both focusing~\cite{Zakharov1974,Han1979,Janssen1983} and
defocusing~\cite{Kuzenstov1988} nonlinear media. In particular, this
kind of symmetry-breaking instability turns a bright-soliton stripe
into a array of two-dimensional filaments~\cite{Mamaev1996} and
bends a dark-soliton stripe creating pairs of optical vortices of
opposite polarities~\cite{Tikhonenko1996}. Consequently, the
transverse instabilities set severe limits on the observation of
one-dimensional spatial solitons in bulk media~\cite{book}.

Several different physical mechanisms for suppressing the soliton
transverse instabilities have been proposed and studied, including
the effect of partial incoherence of
light~\cite{Torres2001,Kristian2003} and anisotropic nonlinear
response~\cite{Kristian2003} in photorefractive crystals, and the
stabilizing action of the nonlinear coupling between the different
modes or polarizations~\cite{Ziad2001}. Recently initiated
theoretical and experimental studies of nonlocal nonlinearities
revealed many novel features in the propagation of spatial solitons
including the suppression of their modulational~\cite{Krolikowski04}
and azimuthal~\cite{Anton06} instabilities. In this Letter we
demonstrate that significant suppression of the soliton transverse
instabilities can be achieved in nonlocal nonlinear media, and we
derive analytical results for the instability growth rate in both
long- and short-scale asymptotic limits.

We consider the propagation of an optical beam in a nonlocal
nonlinear medium described by the normalized two-dimensional
nonlinear Sch{\"o}dinger (NLS) equation,
\begin{equation}
   \begin{array}{l} {\displaystyle
       i \frac{\partial E}{\partial z}+
\frac{1}{2} \Delta_{\perp} E + n E=0,
   } \\*[9pt] {\displaystyle
n-d \Delta_{\perp} n=|E|^2,
   }\end{array}
\label{eq:main}
\end{equation}
where $\Delta_{\perp} =
\partial^2/\partial x^2+\partial^2/\partial y^2$, $E=E(x,y;z)$ is the slowly
varying electric field envelope, $n=n(x,y; z)$ is the optical
refractive index, and the parameter $d$ stands for the strength of
nonlocality. The model (\ref{eq:main}) describes the light
propagation in different types of nonlocal nonlinear media including
nematic liquid crystals~\cite{LC_assanto}.

We look for stationary solutions of Eq.~(\ref{eq:main}) in the form
of the bright soliton stripes, $E(x,y;z)= u(x)\exp(i \beta z)$,
where $u(x)$ is a (numerically found) localized function, $u(\pm
\infty) = 0$, and $\beta$ is the (real) propagation constant.

The transverse instability of quasi-one-dimensional solitons in
nonlocal nonlinear media is investigated by a standard linear
stability analysis~\cite{YSK00}, by introducing the perturbed
solution in the form: $n=n_0(x) + \epsilon\delta n$, and
\[E = e^{i \beta z}[u_0+\epsilon(v+i w)e^{i\lambda z+i p
y}+\epsilon(v^\ast-i w^\ast)e^{-i\lambda^\ast z -i p y}],
\]
where $(u_0, n_0)$ is the solution of Eqs.~(\ref{eq:main}),
$\epsilon\ll1$ is a small perturbation, and $v(x)$, $w(x)$, and
$\delta n(x)$ are perturbed amplitudes that are modulated in the
transverse $y$ direction with the wavenumber $p$. The instability
growth rate is defined as an imaginary part of the eigenvalue
$\lambda$.

Substituting these asymptotic expansions into Eq.~(\ref{eq:main}),
in the first order of $\epsilon$ we obtain a set of linear
equations,

\begin{equation}
   \begin{array}{l} {\displaystyle
       \lambda w = (\beta+\frac{1}{2}p^2) v - \frac{1}{2}\frac{d^2v}{dx^2}-n_0 v-u_0 \delta n,
   } \\*[9pt] {\displaystyle
\lambda v = (\beta+\frac{1}{2}p^2) w - \frac{1}{2}\frac{d^2w}{dx^2}-  n_0 w,
   } \\*[9pt] {\displaystyle
\delta n = d \left(\frac{d^2}{dx^2}-p^2\right) \delta n+2 u_0 v,
   }\end{array}
\label{eq:lin}
\end{equation}
which we then study numerically and analytically.

\begin{figure}[b]
\center
\includegraphics[width=8.3cm]{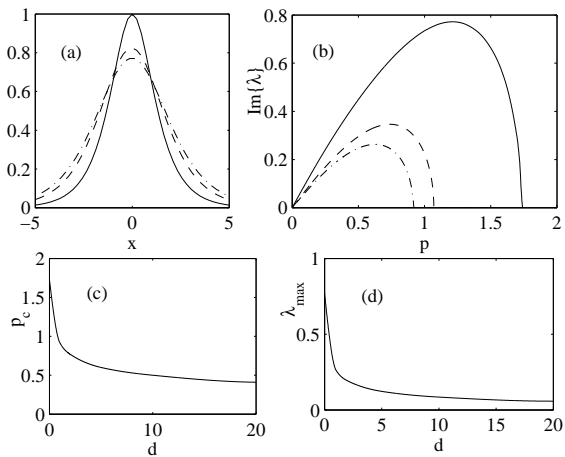}
\caption{(a) Intensity profiles of the bright solitons in local
($d=0$, solid) and nonlocal ($d=0.5$, dashed; $d=1$, dashed-dotted)
nonlinear media. (b) The instability growth rate for bright solitons
vs. the transverse wavenumber $p$ for local (solid) and nonlocal
($d=0.5$, dashed; $d=1$, dash-dotted) nonlinear media. (c,d) Cutoff
value of the transverse wavenumber and the maximum growth rate vs.
the nonlocality parameter $d$.} \label{Fig:F1}
\end{figure}

Typical solutions for bright solitons of the model (\ref{eq:main})
are shown in Fig.~\ref{Fig:F1}(a) for local (solid curve) and
nonlocal (dashed and dashed-dotted curves) media.
Figure~\ref{Fig:F1}(b) shows the growth rate of the soliton
transverse instabilities vs. the modulation wavenumber $p$, for
local ($d=0$) and nonlocal ($d=0.5$ and $d=1$) nonlinearities,
respectively. It can be seen that nonlocality reduces the growth
rate of the transverse instability of the soliton stripe. Moreover,
the cutoff transverse wavenumber $p_c$ of the gain spectrum becomes
smaller as the value of the nonlocality parameter $d$ grows.

To describe the suppression of the soliton transverse instabilities
quantitatively, we calculate the dependence of the cutoff transverse
wavenumber $p_c$ and the maximum growth rate on the strength of
nonlocality, $d$, shown in Figs.~\ref{Fig:F1}(c,d).  We observe that
the maximum growth rate decreases significantly at large values of
the nonlocality parameter, and eventually it approaches zero when $d
\rightarrow \infty$. The cutoff wavenumber $p_c$ of the transverse
instability domain becomes smaller as the value of nonlocality
grows. In the limit of very large values of $d$, the cutoff
wavenumber vanishes as well. Consequently, the soliton stripes tend
to become more stable when the degree of nonlocality increases.

Next, we analyze the transverse instability of bright solitons in
nonlocal nonlinear media by applying a variation method, in accord
with the following steps. First, we expand the nonlocal refractive
index function into series in the nonlocality
parameter~\cite{YYLin2007} in the terms involving the refractive
index in Eqs.~(\ref{eq:lin}). Second, we apply the method of Ref.~3
to construct the asymptotic expansions of the elliptical problem
defined by Eqs.~(\ref{eq:lin}). We employ the corresponding ansatz
defined as: $v=v_0+\Gamma v_1$ and $w=w_0+\Gamma w_1$, where
$\Gamma$ is an imaginary part of the eigenvalue of the linear
stability problem.

\begin{figure}[h]
\center
\includegraphics[width=8.2cm]{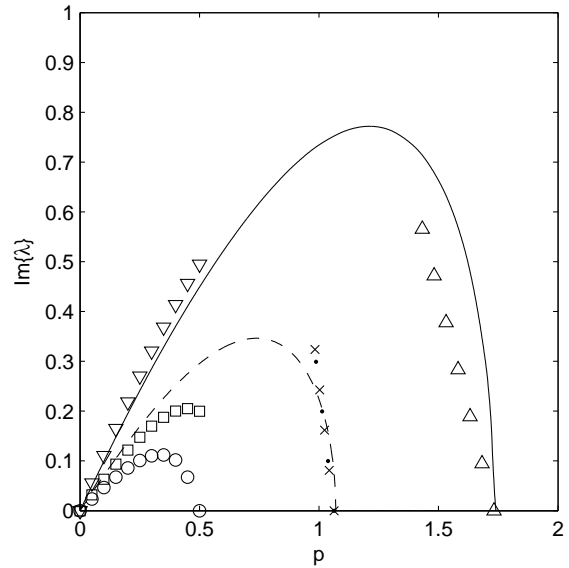}
\caption{Growth rate of the soliton transverse instabilities defined
from numerical simulations and variational approach. Solid and
dashed lines are numerical data for local ($d=0$) and nonlocal
($d=0.5$) nonlinearities. Upper and lower triangular marks: long-scale
and short-scale asymptotic expansions for local nonlinearity.
Squares and circles: long-scale asymptotic expansion based on the
linearized nonlocal eigenvalue system expanded up to the first and second
orders in $d$. Dots and crosses: the short-scale asymptotic expansion
up to the first and second orders in $d$.} \label{Fig:F2}
\end{figure}

In the long-scale expansion, we use the following solutions:
$v_0=0$, $v_1=d^2[{\rm sech}(b_1 x)]/dx^2$, and $w_0=u_0$, where the
parameter $b_1$ is obtained by minimizing the system Lagrangian
corresponding to Eq.~(\ref{eq:lin}),
\begin{eqnarray}
\label{equVA_Lv1} L_{v1}&=&\int_{-\infty}^{\infty} \texttt{d}x
\biggr\{-w_0|v_1|^2+\frac{1}{2}\biggr|\frac{\partial v_1}{\partial
x}\biggr|^2
+\left(\beta+\frac{p^2}{2}\right)|v_1|^2 \nonumber \\
& &-\left(|u_0|^2+d \frac{\partial^2 |u_0|^2}{\partial x^2}\right)|v_1|^2-\frac{2 u_0^2 |v_1|^2}{1+d p^2} \nonumber\\
& &-\frac{2 d}{(1+dp^2)^2}\left(u_0 \frac{\partial^2 u_0}{\partial
x^2}|v_1|^2-u_0^2 \biggr|\frac{\partial v_1}{\partial
x}\biggr|^2\right)\biggr\}.
\end{eqnarray}
Wavenumber $\beta$ and solution $u_0=A {\rm sech}(ax)$ are obtained
by minimizing the Lagrangian of the nonlocal NLS equation based on
the same expansion technique. For a given power $P$ and nonlocality
parameter $d$, we obtain $ \beta = (1/6)Pa+ (1/6)a^2+(2/15)dPa^3$,
$A=(Pa/2)^{1/2}$, and $a= [-5+(25+60dP^2)^{1/2}](12dP)^{-1}$.
While evaluating the system Lagrangian, we use the functional
expansion, ${\rm sec}(ax)\approx {\rm sech}(b_1x)-{\rm
sech}(b_1x){\rm tanh}(b_1x)(a-b_1)x $ to deduce a close form for the
Lagrangian in Eq.~(\ref{equVA_Lv1}). The parameter $b_1$ can
therefore be obtained as a function of both $d$ and $P$. Then, the
instability growth rate is obtained as
\begin{eqnarray}
&&\Gamma^2 = \frac{(w_0,u_0\Delta n(v_1;p=0)) }{( w_0,v_1)}p^2-\frac{1}{4}p^4, \nonumber \\
&&\delta n(v_1;p=0) =  \frac{u_0
v_1}{1+dp^2}+d\frac{\left[\frac{\partial^2}{\partial x^2}(u_0v_1)
\right]}{(1+dp^2)^2},  \label{eq_ldl}
\end{eqnarray}
where we define $(w_0,v_1)=\int_{-\infty}^{\infty}\texttt{d}x
w_0^{\ast} v_1$.

For the asymptotic expansion near the cutoff wavenumber $p_c$, we
use the ansatz $v_0={\rm sech}(b_0x)$, $w_0=0$, and $w_1={\rm
sech}(c_1x)$. Again, to obtain $b_0$ we minimize the Lagrangian
$L_{v_1}$ as a function of the wavenumber $p$. Then the cutoff
wavenumber $p_c$ is found from the condition $L_{v_1}=0$ with an
optimized value of $b_0$. The resulting parameters are deemed as
constant in the Lagrangian for $w_1$,
\begin{eqnarray}
\label{equVA_Lw1} L_{w1}&=&\int_{-\infty}^{\infty}\texttt{d}x
\biggr\{-v_0|w_1|^2+\frac{1}{2}\biggr|\frac{\partial w_1}{\partial
x} \biggr|^2
+ \left(\beta+\frac{p^2}{2}\right)|w_1|^2 \nonumber \\
& &-\left(|u_0|^2+d \frac{\partial^2 |u_0|^2}{\partial
x^2}\right)|w_1|^2\biggr\},
\end{eqnarray}
to deduce parameter $c_1$ as a function of $P$ and $d$. Similarly,
in the short-scale expansion the amplitude growth rate can be
calculated as
\begin{eqnarray}
&&\Gamma^2 = \frac{(v_0,u_0\Delta n(w_1;p=0)) }{( v_0,w_1)}p^2-\frac{1}{4}p^4, \nonumber \\
&&\delta n(w_1;p=0) = \frac{u_0
w_1}{1+dp^2}+d\frac{\left[\frac{\partial^2}{\partial x^2}(u_0w_1)
\right]}{(1+dp^2)^2}. \label{eq_ld2}
\end{eqnarray}

\begin{figure}[t]
\center
\includegraphics[width=8.3cm]{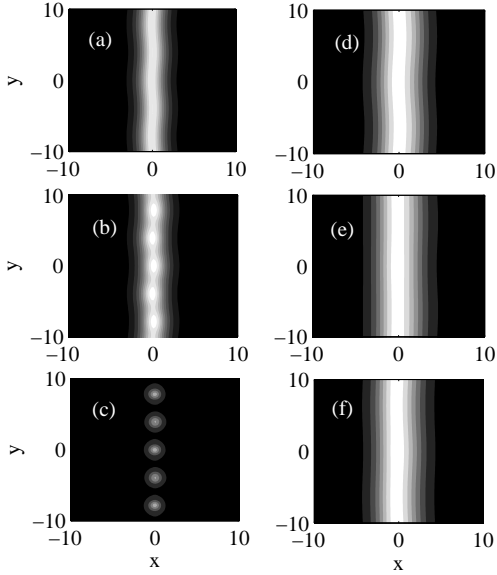}
\caption{Evolution of a modulated bright soliton stripe in (a-c)
local and (d-f) nonlocal nonlinear media at the distances $z=1.0$,
$6.0$ and $10.0$, respectively.} \label{Fig:F3}
\end{figure}

Figure~\ref{Fig:F2} compares the analytical results with the results
of our direct numerical calculations. For nonlocal media, the
long-scale expansion to the first order in $d$ (circles) is
insufficient, but the second-order expansion (squares) gives a good
result. In the short-scale expansion, both the first- (dots) and
second-order (crosses) expansions are in a good agreement with the
numerics.

To observe the consequence of the predicted instability suppression,
we study numerically the evolution of the soliton stripe described
by Eqs.~(\ref{eq:main}). Figures~\ref{Fig:F3}(a-f) compare the
transverse instability of a bright-soliton stripe with the power
density $2$ in local and nonlocal media, with the initial field
modulated transversally with the maximum growth rate. In
Fig.~\ref{Fig:F3}(a-c) we show the snapshots of the soliton stripe
evolution in a local nonlinear medium at $z=1.0$, $6.0$ and $10.0$,
respectively. We observe that at $z=10.0$ [Fig.~\ref{Fig:F3}(c)],
the bright-soliton stripe decays into a sequence of filaments due to
the modulation in the $y$ direction. In a sharp contrast, there is
no visible decay of the soliton stripe in a nonlocal medium ($d=1$)
[Fig.~\ref{Fig:F3}(d-f)], therefore confirming our major conclusion
that the soliton transverse instabilities are suppressed
substantially in nonlocal nonlinear media.

In conclusion, we have analyzed the transverse instabilities of
spatial solitons in nonlocal nonlinear media. By employing the
linear stability analysis and numerical simulations, we have
demonstrated that nonlocal nonlinear response can suppress
significantly the transverse instabilities allowing experimental
observations of the stable propagation of the soliton stripes in
nonlocal media.


\end{document}